\documentclass[twocolumn,prl,showpacs,aps]{revtex4-1}
\usepackage{amssymb}
\usepackage{amsmath}
\usepackage{graphicx}
\usepackage{epsfig}

\begin{document}

\title{An attractor dynamics in a non-Hermitian two-level system}
\author{C. Li, P. Wang}
\author{L. Jin}
\email{jinliang@nankai.edu.cn}
\author{Z. Song}
\email{songtc@nankai.edu.cn}
\affiliation{School of Physics, Nankai University, Tianjin 300071, China}

\begin{abstract}
Exceptional point in non-Hermitian system possesses fascinating
properties. We present an exactly solvable attractor dynamics for the first
time from a two-level time dependent non-Hermitian Hamiltonian. It allows a
way to evolve to the coalescence state from a pure or mixed initial state
through varying the imaginary parameter along a specific diabatic passage.
Contrast to a chaotic attractor that is ultrasensitive to the initial
condition, the designed attractor is insensitive to the initial conditions.
The attractor-like behavior still exists for several adiabatic processes.
\end{abstract}

\maketitle



The non-Hermiticity is described by external imaginary parameters, which can
be imaginary potentials \cite{Bender,klaiman,Guo,Ruter} or nonreciprocal
couplings \cite{LonghiSR}. The non-Hermiticity leads to considerably unusual
features even in simple systems. These include $\mathcal{PT}$ phase
transition~\cite{Bendix,LonghiPRL,Jin1,Scott1,Tony}, unidirectional and
anomalous transport \cite{Kulishov,LonghiOL,Lin,Regensburger,Eichelkraut},
asymmetric reflectionless~\cite{Feng}, and loss induced large nonlinearity
\cite{Peng,Chang}. Exceptional point (EP) is an exclusive critical point in
non-Hermitian systems, at which pairs of eigenstates coalesce and exotic
features occur. Such as invisible defects \cite{LonghiPRA2010,Della,ZXZ},
coherent absorption \cite{Sun} and self sustained emission~\cite%
{Mostafazadeh,LonghiSUS,ZXZSUS,Longhi2015,LXQ}, loss-induced revival of
lasing~\cite{PengScience}, laser-mode selection \cite{FengScience,Hodaei},
as well as $\mathcal{PT}$ chaos \cite{XYL}. In contrast to degenerate
eigenstates, coalesced state is immune from tunneling between coalescence
eigenstates, stabilizing the target quantum state~\cite{LinS}.

The EPs possess fascinating properties, the states in a two-level system
switch when circling an EP after one circle; moreover, the geometric phase
accumulated is circling direction dependent \cite{Dem,Uzdin,Heiss}.
Dynamically encircling an EP was non-adiabatic, the energy transfer is
nonreciprocal \cite{Doppler,Xu,Hassan}. The dissipative dynamics in the
non-Hermitian system were discussed through master equation approach \cite%
{Graefe,Brody}. Arbitrary control of pair polarization was achieved in
complex birefringent metamaterials, the orthogonal polarizations is
generated from nonorthogonal pairs of initial state through dynamical
evolution \cite{Cerjan}.\textbf{\ }These works open up the possibility of
exploring other EPs related dynamical effects. It is interesting to
investigate how a quantum state evolves when a system tends to EP.

In this letter, we propose an exactly solvable time-dependent two-level
system. The system can behave as an attractor when it
tends to or crosses an EP. An attractor can be a point, a curve, or a
surface in the phase space of the system to which orbits are attracted.
Typical observations are an infinite number of unstable orbits embedded in a
chaotic attractor. The chaotic dynamics is sensitive to the initial
conditions, points on two arbitrarily close trajectories may have entirely
distinct dynamics, efforts have been made to obtain improved performance and
multiple uses \cite{Ott}. Nonlinear dissipative system under driven may exhibit 
chaotic behavior \cite{Jeffries}, which were extensively investigated \cite{Grebogi82,Grebogi86,Rosenblum}. 
Recently, periodic or chaotic dynamics were demonstrated for both the optical and mechanical modes caused by the
optomechanical coupling induced nonlinearity \cite{Monifi}. 
Here, an attractor dynamics is presented for the first time from the
two-level non-Hermitian Hamiltonian. The time evolution is studied when the
system approaches its EP. We show that both pure and mixed states evolve to
the coalescence state through varying the imaginary parameter along a
specific diabatic passage. Contrast to a chaotic attractor that is
ultrasensitive to the initial condition, we propose an attractor in a
time-dependent non-Hermitian system without nonlinearity. The attractor is a
limit circle, where the dynamics is insensitive to the initial condition,
the evolution of any state finally converges to one fix orbit. Numerical
simulation shows that the attractor dynamics is applicable to
several adiabatic processes.

\textit{Time-dependent two-level system. }In a Hermitian two-level system,
the transition dynamics is governed by the Landau--Zener formula \cite%
{Landau,Zener,berg,Majorana}, giving the probability of a diabatic
transition between the two energy states. In a non-Hermitian system, EP is
quite different from degenerate states since two eigenstate coalesces into
one eigenstate. Before the construction of a general theory for the dynamics
of the time-dependent system,\ we first present an exact solvable
time-dependent passage.

Any quantum state, being either pure state or mixed state, can be depicted
by a density matrix $\rho =\sum_{i,j}p_{ij}\left\vert i\right\rangle
\left\langle j\right\vert $, where $\left\{ \left\vert i\right\rangle
\right\} $\ denotes a complete orthonormal set, $\langle i\left\vert
j\right\rangle =\delta _{ij}$. For an arbitrary Hamiltonian, including
Hermitian and non-Hermitian ones, the time evolution of the density matrix
obeys the equation%
\begin{equation}
i\frac{\partial }{\partial t}\rho =\left[ H_{+},\rho \right] +\left\{
H_{-},\rho \right\}  \label{se}
\end{equation}%
where we denote $H_{\pm }=(H+H^{\dag })/2$. Here, the square brackets denote
the commutator and the curly brackets denote the anticommutator,
respectively. In principle, the dynamics of a mixed state can be obtained
from the solution of the equation. However, exact analytical solution is
rare, especially for the time-dependent non-Hermitian system with $H(t)\neq
H^{\dag }(t)$.

We consider a simple two-level non-Hermitian system consists of two coupled
cavities $A$ and $B$, the energy is constantly exchanging in the space or
time. The Hamiltonian is%
\begin{equation}
H_{\mathrm{AB}}=\left(
\begin{array}{cc}
i\gamma (t) & \kappa \left( t\right) \\
\kappa \left( t\right) & -i\gamma (t)%
\end{array}%
\right) ,
\end{equation}%
where $\kappa \left( t\right) $ is the strength of the coupling and $\gamma
(t)$\ is the gain or loss of each cavity. All parameters are dependent of
time $t$. The time varying quantities $\gamma (t)$ and $\kappa \left(
t\right) $ satisfy%
\begin{eqnarray}
\gamma (t) &=&\gamma _{n}(t)=t^{2}-2n-1/2,  \label{parameter} \\
\kappa \left( t\right) &=&\kappa _{n}\left( t\right) =1/2-\gamma _{n}(t),
\notag
\end{eqnarray}%
where $n=0,1,2,\cdots $. According to the Eq. (\ref{se}), we have%
\begin{equation}
i\frac{\partial }{\partial t}R\rho =\left[ RH_{\mathrm{AB},+}R^{-1},R\rho %
\right] +\left\{ RH_{\mathrm{AB},-}R^{-1},R\rho \right\}  \label{main}
\end{equation}%
for the density matrix $\rho $ of $H_{\mathrm{AB}}$, where a rotation
transformation
\begin{equation}
R=\frac{1}{\sqrt{2}}\left(
\begin{array}{cc}
1 & i \\
i & 1%
\end{array}%
\right) ,  \label{trans}
\end{equation}%
changes $H_{\mathrm{AB}}$ into%
\begin{equation}
\mathcal{H}_{n}(t)=\left(
\begin{array}{cc}
0 & 1 \\
\omega _{n}^{2}\left( t\right) & 0%
\end{array}%
\right) ,  \label{2x2}
\end{equation}%
Note that $\omega _{n}^{2}\left( t\right) =2n+1-t^{2}$, $(n=0,1,2,...)$, and
$\omega _{n}\left( t\right) $\ can be real or imaginary. For real $\omega
_{n}$, the system has balanced gain and loss. $H_{n}(t)$ becomes a
Jordan-block at $t_{c}=\pm \sqrt{2n+1}$ ($\omega _{n}^{2}\left( t_{c}\right)
=0$) and two eigenvectors coalescence to $\left( 1,0\right) ^{T}$. This is
the EP of the two level system.

Equation (\ref{main}) has exact solution. For the matrix $\mathcal{H}_{n}(t)$%
, the density matrix can be expressed as%
\begin{equation}
\rho _{n}=R\rho =\left(
\begin{array}{cc}
p_{11} & p_{12} \\
p_{21} & p_{22}%
\end{array}%
\right) ,
\end{equation}%
with the elements $p_{ij}$\ satisfying the coupled differential equations%
\begin{equation}
\left\{
\begin{array}{l}
i\dot{p}_{11}=p_{21}-p_{12}, \\
i\dot{p}_{12}=p_{22}-\omega _{n}^{2}p_{11}, \\
i\dot{p}_{21}=\omega _{n}^{2}p_{11}-p_{22}, \\
i\dot{p}_{22}=\omega _{n}^{2}\left( p_{12}-p_{21}\right) .%
\end{array}%
\right.
\end{equation}%
A special solution of above equations is%
\begin{eqnarray}
p_{11} &=&[x_{n}\left( t\right) ]^{2},p_{22}=[y_{n}\left( t\right) ]^{2}=[%
\dot{x}_{n}\left( t\right) ]^{2},  \notag \\
p_{12} &=&(p_{21})^{\ast }=i\dot{x}_{n}\left( t\right) x_{n}\left( t\right)
\end{eqnarray}%
with%
\begin{eqnarray}
x_{n}\left( t\right) &=&\left( 2^{n}n!\sqrt{\pi }\right) ^{-1/2}e^{t^{2}/2}%
\mathrm{h}_{n}\left( t\right)  \notag \\
&=&\left( -1\right) ^{n}\left( 2^{n}n!\sqrt{\pi }\right) ^{-1/2}e^{t^{2}/2}%
\frac{d^{n}}{dt^{n}}e^{-t^{2}}
\end{eqnarray}%
where $h_{n}\left( t\right) $\ is the Hermite polynomial. The solution above
is still valid for the original Hamiltonian $H_{\mathrm{AB}}$ under an
inverse transformation of Eq. (\ref{trans}). The density operator $\rho _{n}$%
\ is the density matrix of the pure state $\left( x_{n},y_{n}\right) ^{T}$,
which is the solution of a quantum harmonic oscillator $\ddot{x}+\omega
_{n}^{2}(t)x=0$. The eigen functions behave distinctly in three different
regions: In the center region, the eigen function is a standing wave inside
and an evanescent wave outside. This solution governs the dynamics when the
system varys through the passage $\omega _{n}^{2}\left( t\right) $, also
provides the exact evolved wave function for a series of initial wave
functions as the system varys along a fixed passage. In other word, for a
given $\omega _{n}^{2}\left( t\right) $, the initial and final states are
two points $\left[ x_{n}\left( t_{\mathrm{1}}\right) ,y_{n}\left( t_{\mathrm{%
1}}\right) \right] ^{T}$ and $\left[ x_{n}\left( t_{\mathrm{2}}\right)
,y_{n}\left( t_{\mathrm{2}}\right) \right] ^{T}$\ on the curve $\left\{
x_{n}\left( t\right) ,y_{n}\left( t\right) \right\} $, respectively, where $%
t_{\mathrm{1}}$\ and $t_{\mathrm{2}}$\ are initial and final instants,
respectively.

\begin{figure}[tbp]
\includegraphics[ bb=0 0 330 285, width=0.35\textwidth, clip]{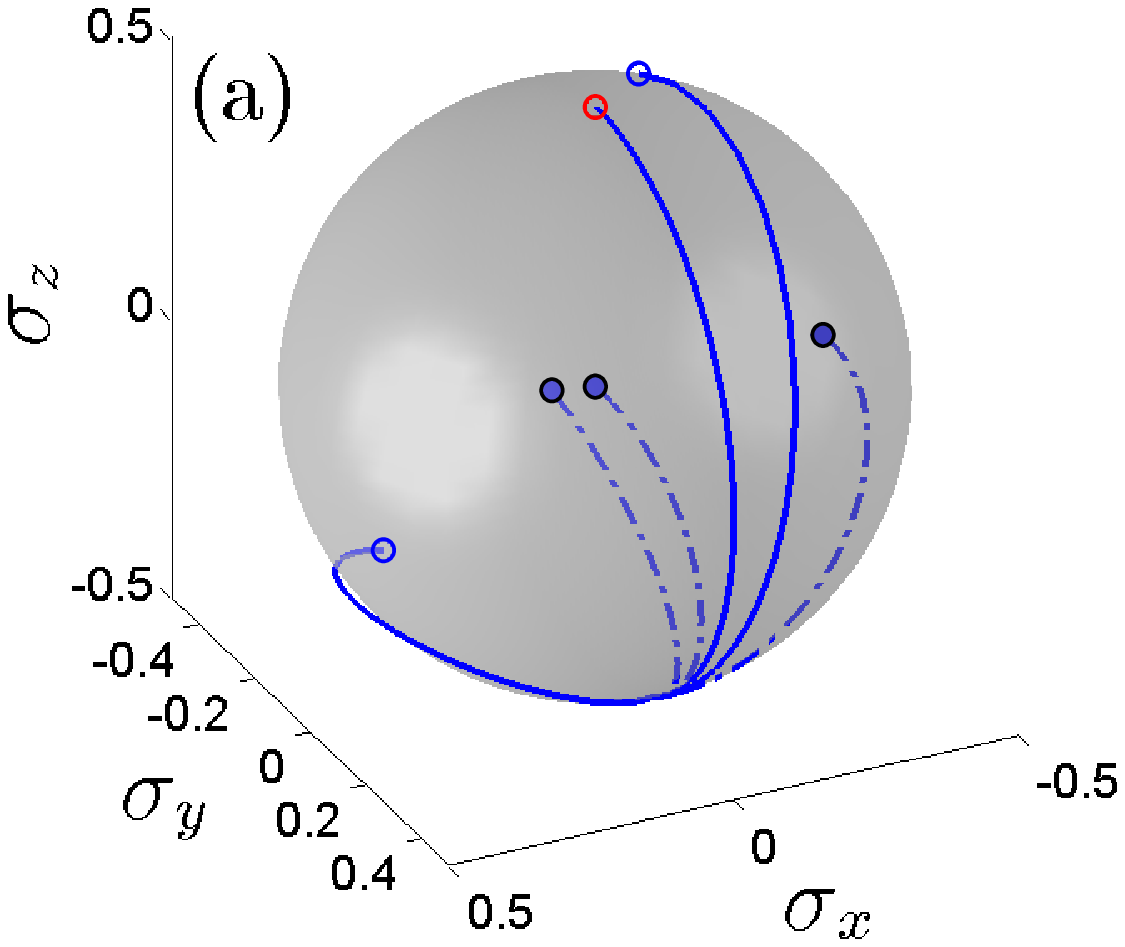} %
\includegraphics[ bb=0 0 330 285, width=0.35\textwidth, clip]{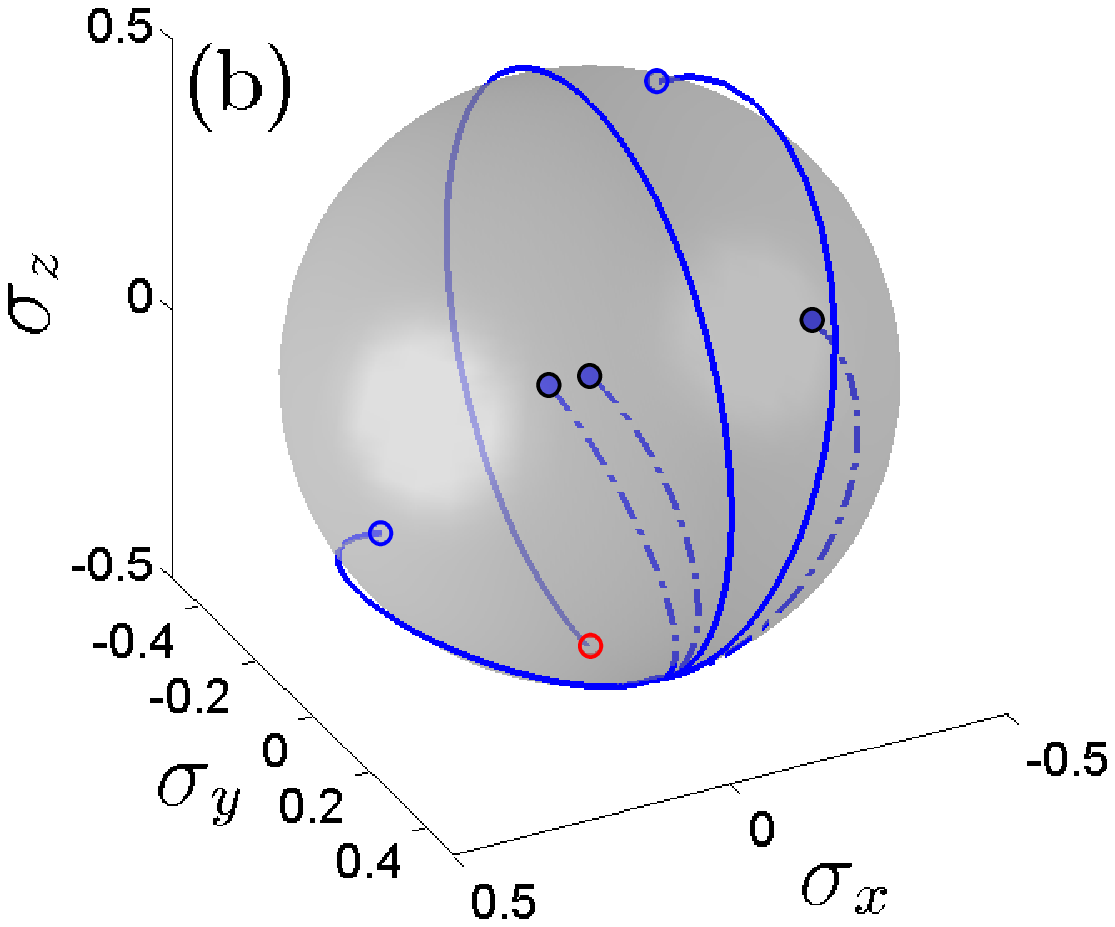}
\caption{(Color online) Effective Bloch dynamics of the non-Hermitian
two-level system (\ref{2x2}) with (a)$n=0$ and (b)$n=1$ from different initial states(including pure and mixed states, which are identified by solid and dash lines). The
red points in the two graphs mean the fixed point $(1,0)$ and $(0,1)$, respectively.}
\label{fig1}
\end{figure}

\textit{Fixed point of evolution. }From the above results, $\left[
x_{n}\left( t\right) ,y_{n}\left( t\right) \right] $ is not the
instantaneous eigenstate of $H_{n}(t)$.\ Thus, the time evolution dynamics
is not a adiabatic process, but a diabatic one. To characterize the process,
we employ the normalized Bloch vector $(\sigma _{x},\sigma _{y},\sigma _{z})$
which is defined as $\sigma _{x}=(x_{n}^{\ast }y_{n}+x_{n}y_{n}^{\ast
})/\Omega $, $\sigma _{y}=-i(x_{n}^{\ast }y_{n}-x_{n}y_{n}^{\ast })/\Omega $%
, $\sigma _{z}=(x_{n}^{2}-\left\vert y_{n}\right\vert ^{2})/\Omega $, with $%
\Omega =2(\left\vert x_{n}\right\vert ^{2}+\left\vert y_{n}\right\vert ^{2})$%
. It directly indicates that $\sigma _{x}=0$\ and $\sigma _{y}^{2}+\sigma
_{z}^{2}=1$, which means that the trajectories of time evolution for any $n$
are the same fixed\ longitude line on the Bloch sphere. This line is an
asymptotic attractor for any nontrivial initial state. The dynamics is
insensitive to the initial states, not only pure states but also mixed
states will evolve to this orbit.

In fact, the solution tells us that $\left[ x_{n}\left( t\right)
,y_{n}\left( t\right) \right] ^{T}\rightarrow \lbrack 0,0]^{T}$\ as $%
t\rightarrow \pm \infty $.\ This indicates that a state probability decays
in the time evolution, similar as a complete absorption; occurring in
certain non-Hermitian systems at the spectral singularity \cite%
{LonghiSUS,ZXZSUS,LXQ,WP,LC}. Inversely, the time-reversal process is
similar as a laser emission.\ However, the probability increasing rate
differs, the probability gain rate is a function of time square. We
demonstrate the features of the dynamics from two limit cases of $n=0$ and $%
n\gg 1$. We show that in both cases, the dynamics is diabatic process and
the state $[1,0]^{T}$ can be dynamically prepared, the preparation
efficiency is $n$ dependent.

When $n=0$, we have%
\begin{equation}
\left\{
\begin{array}{c}
x_{0}\left( t\right) =e^{-t^{2}/2}, \\
y_{0}\left( t\right) =-ite^{-t^{2}/2}.%
\end{array}%
\right.
\end{equation}%
At time $t=t_{\mathrm{c}}=\pm 1$, we have $\omega _{0}^{2}\left( t_{\mathrm{c%
}}\right) =0$ and $\left[ x_{0}\left( t_{\mathrm{c}}\right) ,y_{0}\left( t_{%
\mathrm{c}}\right) \right] ^{T}=e^{-1/2}[1,\mp i]^{T}$, which leads to the
EP of the matrix $H_{0}(t)$. $H_{0}(t)\ $is defective and reduces to a $%
2\times 2$ Jordan block, its instantaneous coalesced eigenstate is $%
[1,0]^{T} $.\ Obviously, the evolved state $\left[ x_{0}\left( t_{\mathrm{c}%
}\right) ,y_{0}\left( t_{\mathrm{c}}\right) \right] ^{T}$ is not the
instantaneous coalesced eigenstate. However, at the instant $t=0$, we have
the evolved state $\left[ x_{0}\left( 0\right) ,y_{0}\left( 0\right) \right]
^{T}=[1,0]^{T}$, which is the coalescing state, while the matrix $%
H_{0}\left( 0\right) =\sigma _{x}$\ is Hermitian with eigenstates $2^{-1/2}%
\left[ 1,\pm 1\right] ^{T}$. If the initial state is $\left[ x_{0}\left(
0\right) ,y_{0}\left( 0\right) \right] ^{T}$ at $t=0$, the final state at $%
t\rightarrow \infty $ approaches a zero vector. On the other hand, there is
an initial state $\left\vert \phi \left( t_{0}\right) \right\rangle =\left(
1+t_{0}^{2}\right) ^{-1/2}\left[ i,t_{0}\right] ^{T}$ at $t_{0}\rightarrow
-\infty $, which can evolve to state $[1,0]^{T}$ at $t=0$ but with infinite
amplitude, the normalized state is $\left\vert \phi \left( 0\right)
\right\rangle /\left\vert \left\vert \phi \left( 0\right) \right\rangle
\right\vert =[1,0]^{T}$. Moreover, any initial state can have this feature
since for an arbitrary state $\left\vert \phi \right\rangle =\left[ \cos
\theta ,e^{-i\phi }\sin \theta \right] ^{T}$, the Dirac inner product $%
\langle \phi \left\vert \phi \left( t_{0}\right) \right\rangle =\left(
1+t_{0}^{2}\right) ^{-1/2}\left( i\cos \theta +t_{0}e^{i\phi }\sin \theta
\right) $ is always nonzero when $t_{0}\rightarrow -\infty $, which
indicates that any state has the component of\ $\left\vert \phi \left(
t_{0}\right) \right\rangle $. This is crucial for the application of an
attractor, i.e., any unknown state evolves to state $[1,0]^{T}$.\ Thus,
robust state preparation via dynamical evolution is possible.

When $n\gg 1$, the range of the oscillating regions can be estimated from
the leftmost or rightmost maximum of the amplitude $x_{n}\left( t\right) $,
where we have%
\begin{equation}
\frac{d}{dt}x_{n}\left( t_{\mathrm{b}}\right) =0,x_{n}\left( t_{\mathrm{b}%
}\right) \approx \pm x_{n-1}\left( t_{\mathrm{b}}\right) ,
\end{equation}%
From the recursion identities%
\begin{equation}
\left\{
\begin{array}{c}
\dot{x}_{n}=\sqrt{\frac{n}{2}}x_{n-1}-\sqrt{\frac{n+1}{2}}x_{n+1}, \\
tx_{n}=\sqrt{\frac{n}{2}}x_{n-1}+\sqrt{\frac{n+1}{2}}x_{n+1},%
\end{array}%
\right.
\end{equation}%
we have%
\begin{equation}
\left\{
\begin{array}{c}
0\approx x_{n-1}-x_{n+1}, \\
\pm t_{\mathrm{b}}x_{n-1}\left( t_{\mathrm{b}}\right) \approx \sqrt{\frac{n}{%
2}}\left( x_{n-1}+x_{n+1}\right) ,%
\end{array}%
\right.
\end{equation}%
which leads to $t_{\mathrm{b}}\approx \pm \sqrt{2n}$. We note that at
instance $t_{\mathrm{b}}$, the solution is the state $[1,0]^{T}$, i.e., $%
x_{n}\left( t_{\mathrm{b}}\right) $ reaches the maximum, while $y_{n}\left(
t_{\mathrm{b}}\right) $\ as the velocity of $x_{n}$\ vanishes.

When we apply the solution to the dynamics of state, similar things happen
as that in $n=0$ case. If the initial state is $[1,0]^{T}$ at $t=t_{\mathrm{b%
}}$, the final state at $t\rightarrow \infty $ approaches zero vector. On
the other hand, the inverse process happens if we take an initial state $%
\left[ x_{n}\left( t_{0}\right) ,y_{n}\left( t_{0}\right) \right] $ at $%
t_{0}\rightarrow -\infty $. It can evolve to state $[1,0]^{T}$ at $t=-t_{%
\mathrm{b}}$ but with infinite amplitude. Comparing to the $n=0$\ case, this
diabatic process is faster. In Fig. \ref{fig1}, we plot the expression of $%
x_{n}\left( t\right) $\ for different $n$ to demonstrate this point.
Similarly, we note that $\omega _{n}^{2}\left( t_{\mathrm{b}}\right) =1$,
while $\omega _{n}^{2}\left( t_{\mathrm{c}}\right) =0$ with $t_{\mathrm{c}}=%
\sqrt{2n+1}$.

So far, we conclude that there exist a series of diabatic passages which can
dynamically prepare the coalescing state $[1,0]^{T}$\ from an unknown
(arbitrary/any) initial state, including mixed state. The duration time
monotonously depends on $n$, i.e., larger $n$ leading to faster process.

To demonstrate our conclusions, numerical simulations are performed for the
evolutions of pure and mixed states. For small time increment $\Delta t$,
the Schr\"{o}dinger equation for density matrix becomes%
\begin{equation}
\rho (t+\Delta t)\approx \rho (t)-i(\left[ H_{+}(t),\rho (t)\right]
-i\left\{ H_{-}(t),\rho (t)\right\} )\Delta t,
\end{equation}%
which is employed to compute the time evolution of density matrix
numerically. A normalized state is depicted by a Bloch vector $\mathbf{a}$,
which is defined as $\rho =\frac{1}{2}(I+\mathbf{a\cdot \sigma )}$, $I$ is
unitary matrix and $\mathbf{\sigma }$\ are Pauli matrices. Therefore, the
trajectory of $\mathbf{a}$\ in the Bloch sphere can describe the time
evolution of a state.

We depict the dynamics of initial states for $n=0$\ and $n=1$\ in Fig. \ref%
{fig1}. From the trajectories, all initial states coincide to the analytical
solution finally. The final state is $[1,0]^{T}$\ for $n=0$\ and $[0,1]^{T}$%
\ for $n=1$. These imply that when $n\in \left( 0,1\right) $, there is a
possibility that we can dynamically prepare any pure state from an unknown
(arbitrary/any) initial state. The numerical results have displayed this
tendency to us.

The dynamics of a $\mathcal{PT}$\ dimer has two fixed points except for the $%
\mathcal{PT}$\ transition point, where two fixed points coincide \cite{Brody}%
. In exact $\mathcal{PT}$\ phase, the orbits are closed circles, which
transform nonorthogonal pair states to orthogonal states using complex
birefringent material \cite{Cerjan}. In broken $\mathcal{PT}$\ phase, the
orbits start from the source fixed point to the sink fixed point. The
attractor in this letter has different dynamics, there exists only one fixed
point, the pure states on the sphere surface and mixed states inside the
sphere both evolve to the fixed point.

\begin{figure}[tbp]
\centering%
\includegraphics[ bb=0 0 330 285, width=0.35\textwidth,
clip]{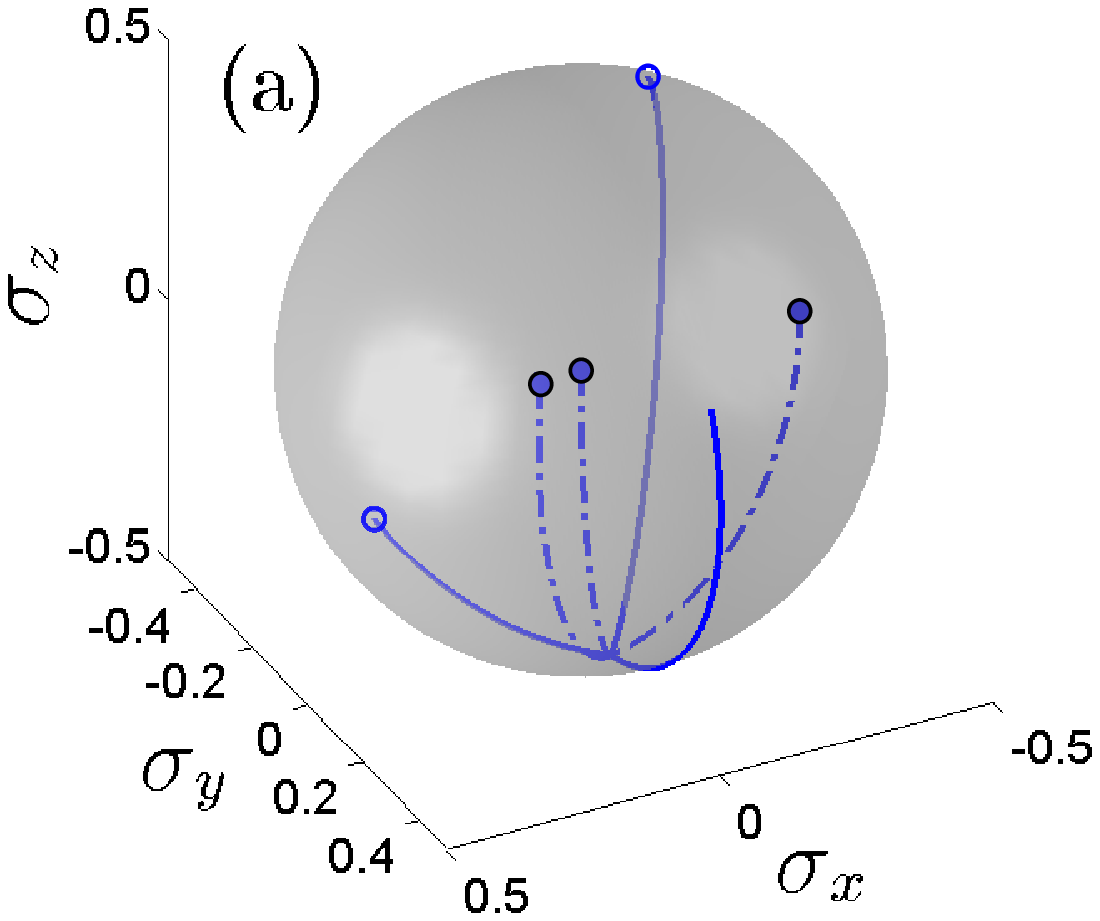}
\includegraphics[ bb=0 0 330 285, width=0.35\textwidth,
clip]{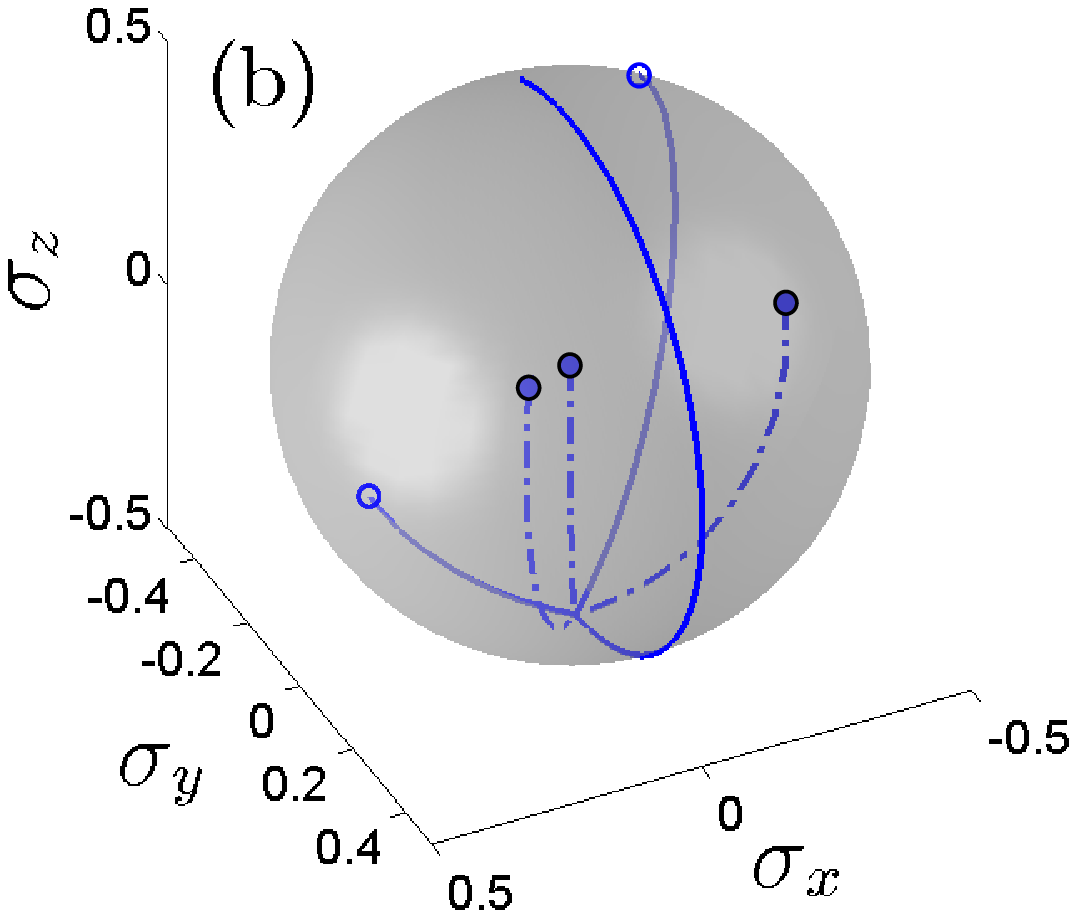} 
\caption{(color online) Effective Bloch dynamics of the non-Hermitian
two-level system: (a)(\ref{sim1}) and (b)(\ref{sim2}) from different initial states
(including pure and mixed states, which are identified by solid and dash lines).}
\label{fig2}
\end{figure}

\textit{Adiabatic process. }Until now, we have proved that the specific
forms of $\gamma (t)$ and $\kappa \left( t\right) $ in $H_{\mathrm{AB}}$ can
lead to the dynamics which has similar phenomena for the complete absorption
and laser emission, i.e., implication of the attractor. So a natural
question is whether an\ adiabatic passage can accomplish the same task. For
example, we consider two passages with $\kappa $ is always a real constant $%
1 $ and (i)%
\begin{equation}
\gamma (t)=t,  \label{sim1}
\end{equation}%
(ii)%
\begin{equation}
\gamma (t)=1-t^{2}.  \label{sim2}
\end{equation}%
Unlike the situation of Hermitian systems, there is no well established
theory to describe the process. On the other hand, we have no idea about the
tunneling between two coalescing levels. Nevertheless, the diabatic solution
implies the possibility of the amplification of amplitude. In this
situation, numerical simulation is a better way to follow a quasi-adiabatic
passage. It is obvious that although the position of EP is different in the
two cases, both of them can go through(or reach) the EP during the time
evolution from a $-t$ satisfied $\kappa ^{2}-\gamma ^{2}(-t)\ll 0$ to $t=0$.
Then according to the simulation, the dynamics of these two cases are same
as what we found before, which implies that once the model can get close to
the EP during the time evolution from the broken area($\kappa ^{2}-\gamma
^{2}(t)\ll 0$ at the beginning of the evolution), there is always a fixed
point in this dynamical process. The simulation result is showed in Fig. \ref%
{fig2}. It indicates that although the speed of the evolution and the
position of the fixed point changed, the asymptotic line, which likes the
one in Fig. \ref{fig1}, still exists even in the adiabatic process with
different forms of $\kappa \left( t\right) $\ and $\gamma \left( t\right) $.

\textit{Experimental realization. }At last we talk something about the
realization of our model. In fact, there is no scheme has been proposed to
realize such a system that both $\kappa \left( t\right) $ and $\gamma \left(
t\right) $ can change with the time in specific forms. But in practice, the
two-level non-Hamiltonian system likes%
\begin{equation}
H=\left(
\begin{array}{cc}
i\gamma (t) & \kappa \\
\kappa & -i\gamma (t)%
\end{array}%
\right)  \label{real}
\end{equation}%
which requires that $\kappa $\ is independent of $t$, has been realized to
study the special feature of the EP point. Such as the emergence of multiple
EPs in the coupled acoustic cavity resonators \cite{Ding} and the properties
associated with encirclement of an EP \cite{Hassan}. In optics, this kind of
system can be also realized by using waveguides or a series of varying
dichroic birefringent plates. We can just let $\gamma (t)=t$ for convenient,
which is also easy to achieve in experiment. Based on the numerical result
before, the dynamics of this system is similar with the original system we
proposed in Eq. (\ref{2x2}). So one can build a simple platform described by
Eq. (\ref{real}) to investigate the phenomena of the attractor, i.e., a
fixed point during the time evolution.

\textit{Summary.} In summary, we have proposed a time-dependent
non-Hermitian two-level system, in which the dynamics has an exactly
solvable passage. By varying the imaginary parameter along a specific
diabatic passage, the two-level system behaves as an attractor, which is
insensitive to the initial conditions. An attractor-like behavior is found
for the first time from a non-Hermitian two-level system. We have shown that
arbitrary pure and mixed state can evolve to the coalescence state.
Moreover, the numerical results indicated two other things: (i) Except the
coalescing state, there is a possibility that we can dynamically prepare
several target states from an unknown (arbitrary/any) initial state. (ii)
The same phenomenon still exists even in adiabatic passages with different
forms of parameters of the cavity. Finally, we give a practical model which
can possibly realize the dynamics of our model in the experiment.

\acknowledgments We acknowledge the support from CNSF (Grant Nos. 11374163
and 11605094) and the Tianjin Natural Science Foundation (Grant No.
16JCYBJC40800).

\end{document}